\documentclass[twocolumn,preprintnumbers,amsmath,amssymb,showpacs]{revtex4}
\usepackage{epsfig}
\usepackage{color}
\usepackage{subfigure}
\usepackage{epstopdf}
\begin{document}
\allowdisplaybreaks
\setlength{\voffset}{1.0cm}
\title{Dimensional reduction in quantum field theories at finite temperature and density}
\author{Johannes Hofmann}
\email{jhofmann@theorie3.physik.uni-erlangen.de}
\affiliation{Institut f\"ur Theoretische Physik III,
Universit\"at Erlangen-N\"urnberg, D-91058 Erlangen, Germany}
\date{\today}
\begin{abstract}
In this work we present two correspondences between the massless Gross-Neveu model with one or two coupling constants in 1+1 dimensions and nonrelativistic field theories in 3+1 dimensions. It is shown that on a mean-field level the massless Gross-Neveu model can be mapped onto BCS theory provided that translational invariance of the condensate is assumed. The generalized massless Gross-Neveu model with two coupling constants is mapped onto a quasi one-dimensional extended Hubbard model used in the description of spin-Peierls systems. It is shown that the particle hole symmetry of the Hubbard model implies self-consistency of the condensate. The dimensional reduction allows an identification of the phase diagrams of the models.
\end{abstract}
\maketitle
\section{Introduction}

Over the last decade, a lot of progress has been made on the solution of toy models belonging to the Gross--Neveu family \cite{1}. They describe $N$ species of massive or massless fermions in $1+1$ dimensions interacting via a scalar or pseudoscalar four--fermion interaction:
\begin{equation}\label{c1}
{\cal L} = \bar \psi \left( {\rm i} \gamma^\mu \partial_\mu - m\right) \psi + \frac{g^2}{2} (\bar \psi \psi)^2 + \frac{G^2}{2} (\bar \psi {\rm i} \gamma^5 \psi)^2 .
\end{equation}
Summation over flavor indices is implied and we abbreviate $\bar \psi \psi \equiv \sum_{i = 1}^N \bar \psi^{(i)} \psi^{(i)}$. For $G^2 \equiv 0$ and $m \equiv 0$ we recover the original Gross--Neveu model (GN) with discrete chiral symmetry $\psi \to \gamma^5 \psi$ \cite{1}. For $g^2 \equiv G^2$ the theory possesses a continuous chiral symmetry $\psi \to e^{{\rm i} \gamma^5 \theta} \psi$ and corresponds to a $1+1$ dimensional version of the Nambu--Jona-Lasinio model ($\chi$GN) \cite{33}. Over the past years, previous work on the phase diagrams of those models \cite{18,32} has been extended significantly and revised phase diagrams were proposed that respect the particle content of the theories. It was found that all of the above models exhibit crystalline phases, for many of which analytical solutions could be obtained \cite{11,58,59,55,60,61}. The GN model features a kink--antikink crystal at high density whereas the thermodynamically preferred ground state of the $\chi$GN model below a transition temperature is a ''chiral spiral'' --- a helical chiral condensate whose amplitude is determined by the temperature and spatial period by the chemical potential \cite{11}. The phase transition is reminiscent of the Peierls transition in 1d metals with the first band gap opening up at the Fermi surface \cite{35}. While those models provide a rich playground for the study of relativistic field theories, they were not thought to have any application to reality due to their lower dimensionality.

In recent years, however, the expertise on those toy models has been used to study problems related to the ground state of QCD at finite density. Thereby, $1+1$ dimensional models arise in the form of effective field theories when phases with lower-dimensional modulations of the chiral condensate are considered. Such modulations can occur due to a strong external field \cite{30} or can be induced by the Fermi surface \cite{34,20}. Examples include work by Shuster and Son \cite{34} who --- based on a dimensional reduction onto a variant of the chirally invariant Thirring model --- refuted the possibility of the large $N_c$ Deryagin--Grigoriev--Rubakov (DGR) chiral wave ground state \cite{27} for $N_c = 3$. More recently, Peierls-like instabilities were discussed in the context of the Quarkyonic Phase of QCD \cite{21} or in the presence of a very strong magnetic field, where they were named ''Quarkyonic Chiral Spirals'' \cite{20, 20a} and ''Chiral Magnetic Spirals'' \cite{30}, respectively. The exact solution of the massive GN model was used by Nickel in order to investigate the phase diagram of the massive Nambu--Jona-Lasinio model when restricting to one-dimensional condensates by mapping the energy spectrum of the model onto the spectrum of the GN model \cite{1b}. Further examples include work by Bietenholz \emph{et al.} \cite{28}.

Since the models in question are very involved, it is desirable to study examples of dimensional reduction isolated from the context of QCD. For this purpose we investigate dimensional reduction in nonrelativistic quantum field theories in this paper. We establish the equivalence of BCS theory of superconductivity to the massless Gross--Neveu model with discrete chiral symmetry on a mean-field level, provided that the chiral condensate does not break translational invariance. Furthermore, we will show that the mean-field description of a well-known model from condensed matter physics which is used in the description of spin-Peierls systems --- the quasi one-dimensional extended Hubbard model at half-filling --- is equivalent to the massless generalized Gross-Neveu model with two coupling constants. In particular, we do not have to impose any restrictions on the symmetry of the condensate and the phase diagrams of both models can be identified.

On the one hand, we aim to supplement the current literature by elucidating the mechanism of dimensional reduction in the context of selected nonrelativistic field theories. On the other hand, we want to explore the fascinating applications of the GN model to condensed matter physics. Furthermore, since Hubbard models are extensively studied both analytically as well as numerically, our results might point to applications of those theories to the GN model.

This paper is structured as follows: In Sec. II we give a brief overview of the basics of the massless GN model with two coupling constants. Section III presents the dimensional reduction of BCS theory onto the GN model with discrete chiral symmetry. The reduction holds only for homogeneous condensates. Limitations for spatially varying condensates as well as the cutoff dependence of the phase diagram are discussed. Section IV contains the dimensional reduction of the quasi one-dimensional Hubbard model onto the generalized GN model. We give an introduction to the model and discuss its particle hole symmetry. Exploiting this symmetry, the dimensional reduction is formulated. A discussion of the cutoff dependence and comparison with selected experimental results follows. The paper is concluded by a summary and conclusions in section V.

\section{Massless Gross--Neveu model with two coupling constants}

In this paper, we consider the massless GN model with two coupling constants (genGN), i.e. as defined by Eq.~(\ref{c1}) with $m \equiv 0$. While early work on this system was carried out by Klimenko \cite{1a}, the full revised phase diagram and particle content were worked out only recently by Boehmer and Thies \cite{2}.

We are interested in the phase diagram of this model. According to Coleman's theorem, any long-range order in 1+1 dimensions is destroyed by fluctuations \cite{25}. As shown by Witten \cite{31}, those fluctuations are suppressed in the limit $N \rightarrow \infty$. The counting of various orders, as carried out in different contexts by Dolan and Jackiw \cite{29} as well as 't~Hooft \cite{3}, reveals that for $N \rightarrow \infty$ a finite leading order can only be obtained if $g^2 N$ and $G^2 N$ are kept fixed. Subleading orders are suppressed by powers of $N$.

The suppression of fluctuations in the large-$N$ limit admits a semiclassical treatment \cite{1,4}. The Hamiltonian of the genGN model in a mean-field approximation becomes
\begin{equation}
\frac{H}{N} = \int {\rm d}x \bigg\{ \psi^\dagger \left[ -{\rm i} \gamma^5 \partial_x + \gamma^0 S + \gamma^1 P \right] \psi + \frac{S^2}{2 g^2 N} + \frac{P^2}{2 G^2 N} \bigg\} ,
\end{equation}
where the scalar and pseudoscalar fields $S$ and $P$ satisfy the self-consistency relations
\begin{equation}\label{b4}
S = - g^2 \langle \bar \psi \psi \rangle \quad {\rm and} \quad P = - G^2 \langle \bar \psi {\rm i} \gamma^5 \psi \rangle
\end{equation}
and $\langle \ldots \rangle$ denotes the thermal expectation value.

The canonical transformation $\psi \to e^{{\rm i} \gamma^5 \pi/4} \psi$ maps the scalar and pseudoscalar interaction terms onto one another. Hence, we can assume without loss of generality that $0 < G^2 \leq g^2$. Renormalization can be performed using the conditions \cite{2}
\begin{equation}
\frac{\pi}{g^2 N} = \ln \left(\sqrt{\left(\Lambda/2\right)^2 + 1} + \Lambda/2\right) \approx \ln \Lambda  \quad {\rm and} \label{a2}
\end{equation}
\begin{equation}
\frac{\pi}{G^2 N} = \ln \Lambda + \xi , \label{a1}
\end{equation}
where we set the scalar condensate equal to 1. $\xi \geq 0$ is the renormalized parameter that describes the imbalance in the scalar and pseudoscalar coupling. The case $\xi \equiv 0$ corresponds to the $\chi$GN model whereas for $\xi \to \infty$ we recover the discrete GN model.

\begin{figure}[b!]
\begin{center}
\scalebox{1.15}{\epsfig{file=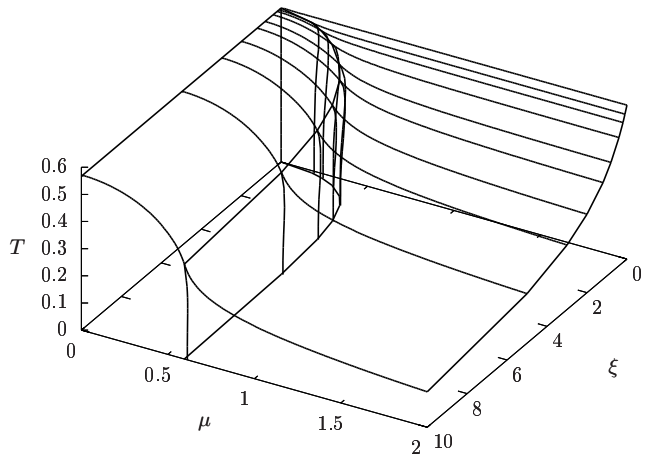}}
\caption{Phase diagram of the genGN model as derived in \cite{2}.} 
\label{fig7}
\end{center}
\end{figure}

The genGN model contains meson and baryon bound states. For a detailed analysis see \cite{4a} and \cite{2}. The phase diagram of the model is shown in Fig.~\ref{fig7}. At zero density (small $\mu$), there is a homogeneous phase with vanishing pseudoscalar condensate. A sheet of second order critical points separates this phase from a massless homogeneous one. Since in both phases the pseudoscalar condensate vanishes, the critical temperature depends on $\mu$ but not on $\xi$. A sheet of first order transition lines divides the massive homogeneous phase from a phase where the condensate takes the form of a soliton crystal which interpolates between the kink-antikink condensate of the GN model and the helical condensate of the $\chi$GN model. All transition sheets converge in a line of tricritical points. The onset of the inhomogeneous phase at zero temperature corresponds to twice the baryon mass. Thereby, at low density and large $\xi$, the condensate takes the form of separated kinks and antikinks and goes over into a sinusoidal shape for large density. As $\xi \to 0$, the pseudoscalar condensate is less suppressed and the ground state oscillates between the scalar and the pseudoscalar condensate. It is interesting to note that chiral symmetry is never restored at zero temperature. In all cases, the spatial period $a$ of the inhomogeneous condensate is determined by the first gap which opens up at the Fermi surface: $a = \pi/p_F$, where $p_F$ is the Fermi momentum. The fact that a system can lower the energy of its ground state by opening up a band gap at the Fermi surface is a well-known phenomenon in condensed matter physics where it was proposed by Peierls for 1d metals \cite{35}.

\section{BCS theory}

\begin{figure}[b!]
\begin{center}
\scalebox{0.75}{\epsfig{file=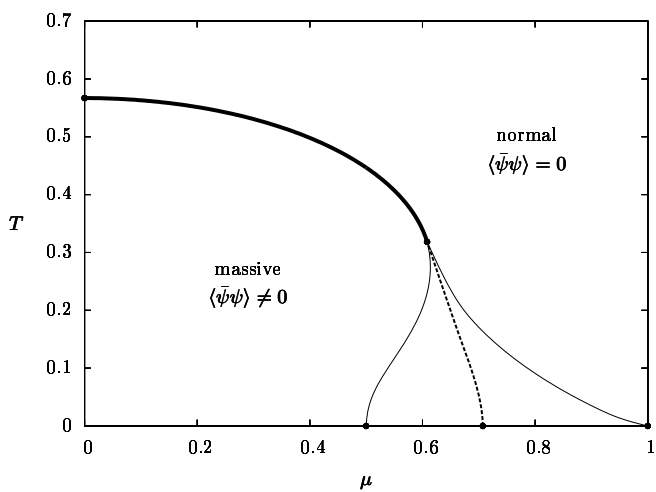}}
\caption{Phase diagram of the massless GN model assuming only translationally invariant phases. The continuous line denotes a second order and the dashed line a first order transition. The thin continuous lines mark boundaries of metastability.} 
\label{fig1}
\end{center}
\end{figure}

We start with the following observation: The phase diagram of the massless GN model was derived for the first time by Wolff \cite{18} assuming only translationally invariant phases. The diagram is shown in Fig.~\ref{fig1}. However, an equivalent phase diagram had been obtained by Sarma \cite{19} even a decade before the proposal of the GN model when investigating the phase diagram of BCS theory in an external magnetic field \cite{26}. The main difference is that the phase diagram of the GN model shows $T$ vs $\mu$ whereas Sarma derives the critical temperature as a function of an external magnetic field. The massive phase of the GN model corresponds to the superconducting phase of BCS theory and the massless phase with restored chiral symmetry to the gapless normal phase. This indicates strongly that both theories are equivalent on that level. This section presents the proof of this equivalence.

The electrons in a conductor interact via the Coulomb interaction with the nuclei and each other. Since the scale of the excitation in a conductor is much smaller than the binding energies due to Coulomb interaction, the conductor can be treated as consisting of a gas of weakly interacting quasiparticles. We start with the effective grand canonical Hamiltonian describing the low-energy excitations in a conductor \cite{24}:
\begin{equation}\label{b1}
H = \int {\rm d}^3{\bf x} \, \psi^\dagger \left( - \frac{\nabla^2}{2 m} - \mu - \mu_A \sigma^3 \right) \psi - \frac{g}{2} (\psi^\dagger \psi)^2 .
\end{equation}
$\psi({\bf x})$ denotes a fermion spinor with components $\psi_\uparrow({\bf x})$ and $\psi_\downarrow({\bf x})$. $m$ is the effective mass of the quasiparticles and $\mu$ the chemical potential. We introduce an ''axial'' chemical potential $\mu_A$ which imbalances the two spin species: The chemical potential of spin-$\uparrow$ fermions becomes $\mu + \mu_A$ and for spin-$\downarrow$ fermions $\mu - \mu_A$. Physically, $\mu_A$ describes an external magnetic field or the effect of impurities in the material. The coupling $g > 0$ carries dimension $E \Lambda^{-3}$, with $E$ carrying dimension of energy and $\Lambda$ of momentum. We perform a Fierz transformation to rewrite the interaction term:
\begin{equation}
(\psi^\dagger \psi)^2 = - \frac{1}{2} \left[\psi^\dagger C \psi^*\right] \left[\psi^T C \psi\right],
\end{equation}
where $C = {\rm i} \sigma_2$. The low-energy excitations are described by states whose momenta ${\bf p}$ are close to the Fermi surface. This defines a cutoff
\begin{equation}\label{b3}
\left| |{\bf p}| - p_F \right| < \Lambda/2 \ll p_F
\end{equation}
with $p_F = \sqrt{2 m \mu}$. In this region, we can linearize the dispersion relation
\begin{equation}
\frac{{\bf p}^2}{2 m} - \mu \approx v_F (|{\bf p}| - p_F).
\end{equation}
The Fermi velocity is defined by $v_F = p_F/m$. All information about the underlying microscopic theory is absorbed in $v_F$ and $\Lambda$. For this linearization to hold for nonzero $\mu_A$ we assume the hierarchy of scales
\begin{equation}
\mu_A \ll v_F \Lambda/2 \ll \mu .
\end{equation}

Renormalization can be performed using a similar condition as Eq.~(\ref{a2}) \cite{61a}:
\begin{equation}\label{b2}
\frac{2}{g \rho} = \ln v_F \Lambda,
\end{equation}
where we set the energy scale equal to $1$ and define the density of states \cite{62}
\begin{equation}
\rho = \frac{p_F^2}{\pi^2 v_F}.
\end{equation}
In deriving Eq.~(\ref{b2}) we apply a large $N$ argument: Since we only consider states close to the Fermi surface, the fermion propagator $G({\bf p})$ has nonvanishing support in a shell around the Fermi surface defined by Eq.~(\ref{b3}). This means that for $|{\bf q}| \sim {\cal O}(p_F)$ the product of two propagators $\int {\rm d}{\bf p} \, G({\bf p}) G({\bf q + p})$ is phase-space suppressed by a factor of ${\cal O}(\Lambda/p_F)$. To leading order, this selects all ''cactus'' or ''daisy'' diagrams. For further reference see \cite{22,29}.

By virtue of the above argument we analyze (\ref{b1}) in a mean-field approximation. Eq. (\ref{b1}) then becomes
\begin{align}\label{b4}
H &= \int {\rm d}^3{\bf x} \ \huge \bigg\{ \psi^\dagger \left(v_F (|- {\rm i} \nabla| - p_F) - \mu_A \sigma_3\right) \psi
\nonumber \\
&\qquad + \frac{1}{2} \Delta \psi^\dagger C \psi^* - \frac{1}{2} \Delta^* \psi^T C \psi + \frac{\Delta^2}{g}\bigg\},
\end{align}
where the BCS-condensate $\Delta$ satisfies the self-consistency condition
\begin{equation}
\Delta = \frac{g}{2} \langle \psi^T C \psi \rangle
\end{equation}
and $\langle \ldots \rangle$ denotes the thermal average.

\subsection{Dimensional Reduction}

To keep track of the dimensionality of the various quantities we place the system in a box of length $L$. We define the Fourier transform of the spinors by
\begin{equation}
\psi({\bf x}) = \frac{1}{L^{3/2}} \sum_{\bf p} \psi({\bf p}) e^{{\rm i} {\bf p} \cdot {\bf x}} .
\end{equation}
Since the Hamiltonian is rotationally invariant, we can switch to spherical coordinates and formally replace the angular integration by the sum over a number of $N_{\rm pat}$ patches that cover the Fermi surface:
\begin{equation}\label{b5}
\int \frac{{\rm d}^3{\bf p}}{(2 \pi)^3} \to N_{\rm F} \left[\frac{1}{N_{\rm pat}} \sum_i\right] \frac{1}{L^3} \sum_{\rm p} ,
\end{equation}
where we set $N_{\rm F} = p_F^2 L^2/\pi = 4 \pi p_F^2/(2 \pi/L)^2$. Note that this is the number of patches that cover the Fermi surface if the system is enclosed in a finite box of length $L$, hence $N_{\rm pat} \equiv  N_{\rm F}$. We introduce the notation $\psi^{(i)}(p)$ for a spinor with momentum $p_F + p$ in the direction of the $i$-th patch. We label the patch opposite to a patch $i$ by $-i$. Eq.~(\ref{b4}) becomes (suppressing the patch labels)
\begin{align}
\frac{H}{N_{\rm pat}} &= \sum_{\rm p} \ \huge \left\{ \psi^\dagger \left(v_F p - \mu_A \sigma_3\right) \psi + \frac{1}{2} \Delta \psi^\dagger C \psi^* \right. \nonumber \\
&\qquad \left. - \frac{1}{2} \Delta \psi^T C \psi \right\} + \frac{L \Delta^2/v_F}{\pi \rho g} .
\end{align}
The factor  $g \rho$ is dimensionless. If we perform a partial particle-hole conjugation (not changing the spin quantum number) for spin-$\downarrow$ particles,
\begin{equation}
\left(\begin{matrix}\psi^{(i)}_\uparrow(p) \\ \psi^{(i)}_\downarrow(p)\end{matrix}\right) \rightarrow \left(\begin{matrix}\psi^{(i)}_\uparrow(p) \\ \psi^{(- i) \dagger}_\downarrow(p)\end{matrix}\right) ,
\end{equation}
we obtain after normal ordering the Hamiltonian of the massless Gross--Neveu model in a chiral basis:
\begin{align}
\frac{H}{N_{\rm pat}} &= \sum_{\rm p} \ \psi^\dagger \left(v_F p \, \sigma_3 - \mu_A + \Delta \, \sigma_1\right) \psi + \frac{L \Delta^2/v_F}{\pi g \rho} \nonumber \\
&= \sum_{\rm p} \ \bar{\psi} \left(v_F p \, \gamma^1 - \mu_A \, \gamma^0 + \Delta\right) \psi + \frac{L \Delta^2/v_F}{\pi g \rho}, \label{d1}
\end{align}
where we choose $\gamma^0 = \sigma_1, \gamma^2 = -{\rm i}\sigma_2$ and $\gamma^5 = \sigma_3$. This establishes the equivalence between both models on a mean-field level. Each flavor in the GN model corresponds to a patch on the Fermi sphere. Spin-$\uparrow$ and -$\downarrow$ are mapped onto the right- and left-handed components of the relativistic spinor. The coupling $g \rho$ corresponds to $(2/\pi) \, g^2 N$ in the Gross--Neveu model.

\begin{figure}[t!]
\begin{center}
\scalebox{1.09}{\epsfig{file=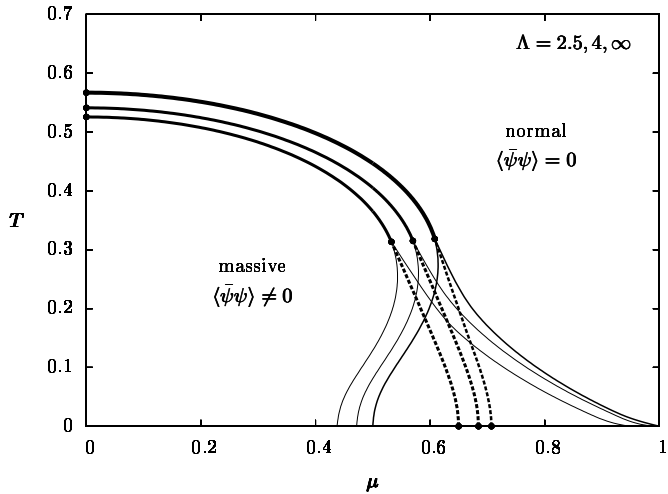}}
\caption{Phase diagram of the massless GN model assuming only translationally invariant phases for different cutoff values (inner to outer) $\Lambda = 2.5, 4$ and $\infty$. The thin continuous lines mark boundaries of metastability.}
\label{fig2}
\end{center}
\end{figure}

Unlike in a relativistic field theory, the cutoff is a physical quantity in BCS theory. Typical cutoffs are of order ${\cal O}(10^2)$--${\cal O}(10^3)$ in units of the mass scale \cite{48a}. In order to establish the full equivalence we must assure that for typical values of the cutoff the phase diagram of the theory is not significantly distorted and we can take $\Lambda \to \infty$ without loss of generality. The Hamiltonian (\ref{d1}) is just the Hamiltonian of a massive free relativistic Fermi gas with single particle energies $\varepsilon(p) = \pm \sqrt{p^2 + \Delta^2}$ and chemical potential $\mu_A$ plus a \emph{c}--number term. The grand canonical potential density is given by (setting $v_F \equiv 1$)
\begin{align}
\frac{\Omega}{N L} &=  \int_{- \Lambda/2}^{\Lambda/2} \frac{{\rm d}p}{2 \pi} \ln \left\{ \left(1 + e^{\beta (\varepsilon(p) + \mu_A)}\right) \left(1 + e^{- \beta (\varepsilon(p) - \mu_A)}\right) \right\} \nonumber \\
&\qquad + \frac{\Delta^2}{2 g^2 N} .
\end{align}
Removing the logarithmic divergences by using the renormalization condition Eq.~(\ref{a1}) (or Eq.~(\ref{b2})) without taking the limit $\Lambda \to \infty$ we obtain
\begin{align}
\frac{\Omega}{N L} &= - \frac{2}{\beta} \int_0^{\Lambda/2} \frac{{\rm d}p}{2 \pi} \ln \left\{ \left(1 + e^{- \beta \sqrt{p^2 + \Delta^2} - \mu_A}\right) \right. \nonumber \\
&\quad \times \left. \left(1 + e^{- \beta \sqrt{p^2 + \Delta^2} + \mu_A}\right) \right\} + \frac{\Delta^2}{2 \pi} \left(\ln \Delta - \frac{1}{2}\right) \nonumber \nonumber \\
&\quad + \frac{\Delta^2}{2 \pi} \ln\left(\frac{\Lambda}{2} + \sqrt{\left(\frac{\Lambda}{2}\right)^2 + 1}\right) \nonumber \\
&\quad - \frac{\Delta^2}{2 \pi} \ln\left(\frac{\Lambda}{2} + \sqrt{\left(\frac{\Lambda}{2}\right)^2 + \Delta^2}\right) \nonumber \\
&\quad + \frac{1}{16 \pi} \frac{\Delta^4}{(\Lambda/2)^2} + \mathcal{O}\left(\Delta^2 \left(\frac{\Delta}{\Lambda/2}\right)^4\right) , \label{f1}
\end{align}
where we subtracted two trivial ''would-be'' divergences
\begin{equation}
- \frac{1}{8 \pi} \Lambda^2 - \frac{\mu_A}{2 \pi} \Lambda ,
\end{equation}
the second one stemming from the infinite fermion density of the Dirac sea. The second line of Eq.~(\ref{f1}) vanishes in the limit $\Lambda \to \infty$. Fig.~\ref{fig2} shows several phase diagrams for cutoff values $\Lambda = 2.5,4$ and $\infty$. As we can see, we can take $\Lambda \to \infty$ for typical values of the cutoff.

It is not possible, however, to include inhomogeneous phases of superconductors \cite{49,50} in the dimensional reduction in the same vein as sketched above. Consider the Hamiltonian Eq.~(\ref{b4}) for a spherically symmetric condensate,
\begin{align}
&\frac{1}{2} \sum_{\bf p} \sum_{\bf q} \Delta(|{\bf q}|) \psi^\dagger(|{\bf p + q}|) C \psi^*(|{\bf p}|) \nonumber \\
&\qquad - \frac{1}{2} \Delta^*(|{\bf q}|) \psi^T(|{\bf p - q}|) C \psi(|{\bf p}|) ,
\end{align}
where $|{\bf q}| \ll |{\bf p}|$. The requirement for dimensional reduction is
\begin{equation}
|{\bf p \pm q}| = |{\bf p}| \pm \frac{{\bf p} \cdot {\bf q}}{|{\bf p}|} = |{\bf p}| \pm |{\bf q}|
\end{equation}
for all ${\bf p}$. This is only possible for ${\bf q} = 0$, i.e. a homogeneous condensate.

\section{Hubbard model}

In this section we will consider a system from condensed matter physics that displays a one-dimensional instability. For a wide class of organic and some inorganic materials conduction is essentially restricted to one dimension due to their anisotropic structure. They can be described as a family of weakly coupled chains. This weak transverse coupling allows these materials to circumvent the Coleman--Mermin--Wagner theorem and exhibit long-range order. Such phases are characterized by a one-dimensional inhomogeneous charge or spin distribution and are, therefore, called charge and spin density waves (CDW, SDW). These materials are known as spin-Peierls systems. For further reference see \cite{44,63}.

As for BCS theory, the phase diagram of the Hubbard model does not depend on the chemical potential $\mu$ but on an external magnetic field $h$. At low temperature and small magnetic field the system possesses a CDW${}_0$ ground state. The CDW${}_0$ condensate takes the form of a plane wave $\Delta e^{{\rm i} {\bf Q}_n \cdot {\bf x}}$. The wavevector ${\bf Q}_n$ is tilted by an angle depending on the lattice spacing with respect to the preferential direction. At high density, this condensate is modulated in $x$ direction. This is called a CDW${}_x$ phase. The modulation is of order $\cal{O}\left( {\bf Q_n} \right)$. A CDW${}_y$ condensate which is modulated in the perpendicular directions is possible as well, but is excluded for the parameter values that we consider in this paper \cite{39}. At high temperature a transition to a homogeneous normal phase occurs where the gap vanishes. In this section we will show that the phase diagram of the Hubbard model can be identified with the phase diagram of the genGN model, Fig.~\ref{fig1}. The massive homogeneous phase in the genGN model corresponds to a CDW${}_0$ phase, the inhomogeneous phase to a CDW${}_x$ phase and the massless chirally symmetric phase to the normal phase.

Early work on the relationship between the original Hubbard model \cite{38} with repulsive interaction and the chirally invariant Thirring model was carried out by Filev \cite{23} and extended by Melzer \cite{36} as well as Woynarovich and Forg\'acs \cite{37}. Since we are investigating a more general model which contains two competing interaction terms and admits a nontrivial phase diagram \cite{2}, there is almost no overlap between our work and theirs.

\subsection{Definition, Symmetries and Anisotropic Hopping}

We start with a system of fermions on a hypercubic lattice with $N_{\rm lat}$ sites that are allowed to tunnel (or ''hop'') to nearest lattice sites and to interact with their nearest neighbors via a spin-dependent repulsive quartic interaction. The lattice is bipartite, i.e. it can be divided into two sublattices A and B such that the interaction takes place between fermions on different sublattices:
\begin{align}\label{v1}
H &= - \sum_\sigma \sum_{<{\bf i j}>} t_{ij} \psi_\sigma^\dagger({\bf j}) \psi_\sigma({\bf i}) - h \sum_\sigma \sum_{\bf j} \sigma \, \psi_\sigma^\dagger({\bf j}) \psi_\sigma({\bf j}) \nonumber \\
&\quad - \frac{1}{2} \sum_{\sigma, \sigma'} \frac{1}{N_\mu} \sum_{<{\bf i j}>} V_{\sigma \sigma'} \,  \Big(\psi_\sigma^\dagger({\bf i}) \psi_\sigma({\bf i}) - n_{\sigma}({\bf i})\Big) \nonumber \\
&\quad \times \Big(\psi_{\sigma'}^\dagger({\bf j}) \psi_{\sigma'}({\bf j}) - n_{\sigma'}({\bf j})\Big) ,
\end{align}
The sum runs over the spin indices $\sigma$ and lattice sites ${\bf i}$. $<{\bf i j}>$ denotes the sum over nearest neighbors and $N_\mu$ the number of nearest neighbors ($N_\mu = 2 d$ for a hypercubic lattice). $\psi_\sigma({\bf i})$ is a fermion spinor at site ${\bf i}$ with spin $\sigma$ and $n_\sigma({\bf j}) = \langle \psi_\sigma^\dagger({\bf j}) \psi_\sigma({\bf j}) \rangle$ is the mean occupation number of a site. $t_{ij}$ is the hopping amplitude between adjacent lattice sites which we suppose to depend only on the lattice direction. As for the BCS theory, $h$ describes the effect of impurities or an external magnetic field. The most general symmetric form of the coupling is
\begin{equation}
V_{\sigma \sigma'} = U_c - U_s \sigma \sigma' 
\end{equation}
with $U_c < 0$ and $0 \leq U_s \leq |U_c|$. A coupling of the form $- U_s (1 + \sigma \sigma')$ is repulsive for spins of the same type and favors the formation of an alternating pattern of spin-$\uparrow$ and -$\downarrow$ fermions --- a spin density wave (SDW). A coupling of the form $U_c < 0$ enhances an inhomogeneous charge distribution --- a charge density wave (CDW).

The Hamiltonian (\ref{v1}) is symmetric under a particle hole conjugation defined by
\begin{align}
U^\dagger \psi_\sigma({\bf j}) U &= \left\{\begin{matrix} \psi_{-\sigma}^\dagger({\bf j}) & {\bf j} \in A \\ - \psi_{-\sigma}^\dagger({\bf j}) & {\bf j} \in B\end{matrix}\right. = e^{{\rm i} {\bf Q}_n \cdot {\bf x}_j} \, \psi_{-\sigma}^\dagger({\bf j}) ,
\end{align}
where ${\bf Q}_n = (\pi/a_x, \pi/a_y, \pi/a_z)$ and ${\bf x}_j = (j_1 a_x, j_2 a_y, j_3 a_z)$ and $a_k$ is the lattice spacing in $k$ direction. The minus sign ensures that the kinetic term is invariant under this transformation for a bipartite lattice and the factor $\sigma$ in the magnetic term provides the invariance of this term. The invariance of the interaction term follows from $U^\dagger \psi_\sigma^\dagger({\bf i}) \psi_\sigma({\bf i}) U = 1 - \psi_{-\sigma}^\dagger(\bf{i}) \psi_{-\sigma}(\bf{i})$ and the symmetry in the summation over $\sigma$ and $\sigma'$ as well as $<\bf{i j}>$. The momentum decomposition of the transformed fields is
\begin{align}\label{v4}
U^\dagger \psi_\sigma({\bf p}) U &= \frac{1}{N_{\rm lat}^{1/2}} \sum_{\bf j} U^\dagger \psi_\sigma({\bf j}) U \, e^{-{\rm i} {\bf p} \cdot {\bf x}_j} \nonumber \\
&= \psi_{-\sigma}^\dagger({\bf Q}_n - {\bf p}) .
\end{align}
Note that the transformation reflects the momentum on the Fermi surface. For further reference, we note an identity that holds for the expectation value of operator bilinears (when ${\bf q} \neq 0$) \cite{37a}:
\begin{align}
\langle \psi_\sigma^\dagger({\bf p} - {\bf q}) \, \psi_\sigma({\bf p}) \rangle &= \frac{{\rm tr}\left\{\psi_\sigma^\dagger({\bf p} - {\bf q}) \psi_\sigma({\bf p}) \exp\left[- \beta H\right] \right\}}{{\rm tr} \left\{\exp\left[- \beta H\right]\right\}} \nonumber \\
&= - \langle \psi_{-\sigma}^\dagger({\bf Q}_n - {\bf p}) \, \psi_{-\sigma}({\bf Q}_n - {\bf p}+ {\bf q}) \rangle.
\end{align}
For ${\bf q} = 0$ the same calculation yields the expression $\langle \psi_\sigma^\dagger({\bf j}) \psi_\sigma({\bf j}) \rangle + \langle \psi_{-\sigma}^\dagger({\bf j}) \psi_{-\sigma}({\bf j}) \rangle = 1$. This symmetry implies half-filling of the system, i.e. half of the number of states are occupied:
\begin{equation}
\sum_\sigma \sum_{\bf j} \langle \psi_\sigma^\dagger({\bf j}) \psi_\sigma({\bf j}) \rangle = N_{\rm lat} .
\end{equation}

The low-energy behavior is determined by the modes close to the Fermi surface. For ${\bf Q} \approx 0$ only a fluctuation term survives in Eq.~(\ref{v1}) which we will ignore. It is customary to neglect scattering into higher states and assume ${\bf Q} \approx \pm {\bf Q}_n$. Hence, we obtain the Hamiltonian
\begin{align}\label{v5}
H &= \sum_\sigma \sum_{\bf p} \varepsilon_\sigma({\bf p}) \psi_\sigma^\dagger({\bf p}) \psi_\sigma({\bf p}) + \frac{1}{2} \sum_{\sigma, \sigma'} \sum_{{\bf p}, {\bf p}', {\bf Q}} V_{\sigma \sigma'} \psi_\sigma^\dagger({\bf p} + {\bf Q}) \nonumber \\
&\quad \times \psi_\sigma({\bf p}) \psi_{\sigma'}^\dagger({\bf p}' - {\bf Q}) \psi_{\sigma'}({\bf p}') ,
\end{align}
where the single particle energy-spectrum $\varepsilon_\sigma({\bf p})$ is given by
\begin{equation}
\varepsilon_\sigma({\bf p}) = \varepsilon({\bf p}) - \sigma h = - \sum_i 2 t_i \cos p_i a_i - \sigma h .
\end{equation}
We note that this Hamiltonian is still symmetric under a particle hole transformation. The spin-independent part of the energy is antisymmetric under a shift by ${\bf Q}_n$:
\begin{equation}\label{v3}
\varepsilon({\bf p} + {\bf Q}_n) = - \varepsilon({\bf p}) .
\end{equation}
In particular, every point on the Fermi surface of the free field theory at half-filling and $h = 0$ is mapped onto another point on the Fermi surface under a shift by ${\bf Q}_n$. This is called nesting and ${\bf Q}_n$ is known as the nesting vector.

\subsection{Dimensional Reduction}

In a quasi one-dimensional system the hopping amplitude in one direction is much larger compared to the others: $t_x \gg t_y, t_z$. If we assume that the hopping amplitude is proportional to the overlap of atomic orbitals this reflects directly the orbital structure and the configuration of the lattice. A slice through the Fermi surface is shown in Fig.~\ref{fig3}. We linearize the energy around $p_F$:
\begin{figure}[b!]
\begin{center}
\scalebox{0.75}{\epsfig{file=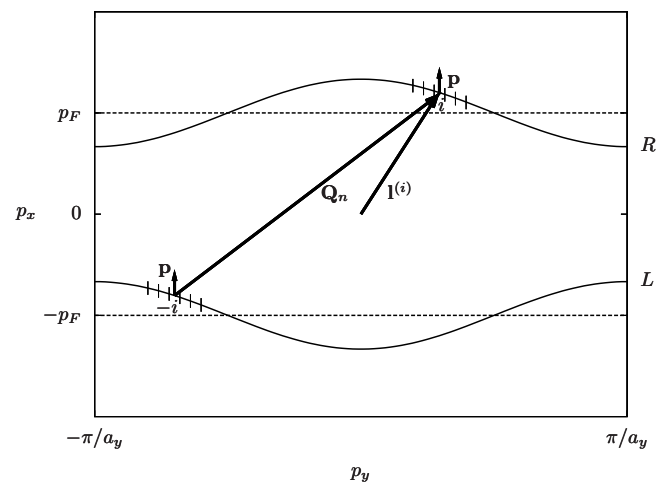}}
\caption{Slice through the Fermi surface of the quasi one-dimensional Hubbard model.}
\label{fig3}
\end{center}
\end{figure}
\begin{equation}
\varepsilon_\sigma({\bf p}) \approx v_F (p - p_F) - 2 t_y \cos p_y a_y - 2 t_z \cos p_z a_z - \sigma h,
\end{equation}
where the Fermi velocity is defined by
\begin{equation}
v_F = \left. \frac{\partial \varepsilon({\bf p})}{\partial p_x} \right|_{p_x = \pi/2 a_x,p_y = p_z = 0} = 2 t_x a_x .
\end{equation}
Again we assume the hierarchy $h,t_y,t_z \ll v_F p_F$.

We analyze the Hamiltonian in a mean-field approximation
\begin{align}\label{v5a}
H &= \sum_{{\bf p}, \sigma} \varepsilon_\sigma({\bf p}) \psi_\sigma^\dagger({\bf p}) \psi_\sigma({\bf p}) + \sum_{{\bf Q}, {\bf p}, \sigma} \psi_\sigma^\dagger({\bf p} + {\bf Q}) \psi_\sigma({\bf p}) \Delta_{{\bf Q} \sigma} \nonumber \\
&\quad - \frac{1}{2} \sum_{{\bf Q}, \sigma}  D_{- {\bf Q} \sigma} \Delta_{{\bf Q} \sigma} ,
\end{align}
where we define
\begin{align}
D_{{\bf Q} \sigma} &= \sum_{\bf p} \langle \psi_{\sigma}^\dagger({\bf p} - {\bf Q}) \psi_{\sigma}({\bf p}) \rangle \quad {\rm and} \nonumber \\
\Delta_{{\bf Q} \sigma'} &= \sum_{\sigma'} V_{\sigma \sigma'} D_{{\bf Q} \sigma'} .
\end{align}

Because of the nesting symmetry (\ref{v3}) the Fermi surfaces for $p_x > 0$ and $p_x < 0$ can be mapped onto one another for small perturbations. In this case, each $p_x$ value corresponds to two patches on the Fermi surface with $p_x > 0$ and $p_x < 0$, respectively. For the right (upper) Fermi surface (i.e. $p_x > 0$) we introduce the notation $\psi_{R\sigma}^{(i)}(p) = \psi_\sigma({\bf l}^{(i)} + (p,{\bf 0}_\perp))$, where ${\bf l}^{(i)}$ points to the i-th patch on the right Fermi surface. If we define left-moving spinors by $\psi_{L\sigma}^{(i)}(p) = \psi_\sigma({\bf l}^{(i)} - {\bf Q}_n + (p,{\bf 0}_\perp))$ the mean-field Hamiltonian separates into a sum over all patches. The construction is illustrated in Fig. \ref{fig3}. In this notation the action of the particle-hole transformation (\ref{v4}) on the spinors reads
\begin{equation}
U \psi_{L/R \sigma}(p) U^\dagger = \psi_{L/R -\sigma}^\dagger(-p) .
\end{equation}
Note that by definition $D_{{\bf Q} \sigma}^\dagger = D_{- {\bf Q} \sigma}$ and by the particle-hole symmetry (\ref{v4}) 
\begin{equation}\label{v6}
D_{{\bf Q}_n + q \sigma} = D_{- {\bf Q}_n + q -\sigma} .
\end{equation}

For our choice of couplings $U_c < 0$ and $|U_c| \geq U_s > 0$ it can be shown that the condensate $\Delta$ is only modulated in $x$ direction \cite{39}. Such modulations are called CDW${}_x$ phases (as opposed to CDW${}_y$ phases, where the modulation is perpendicular to the conducting direction). We can now split the summation over ${\bf p}$ into a sum over patches on the Fermi sphere and a summation over the $p_x$-component. Equation~(\ref{v5a}) becomes
\begin{widetext}
\begin{align}\label{v7}
\frac{H}{N} &= \sum_\sigma \sum_{p, q} \left(\psi_{R\sigma}^\dagger(p + q) \ \psi_{L\sigma}^\dagger(p)\right) \left(\begin{matrix} \left[v_F (p + q) - \sigma h\right] \delta_{q,0} & \Delta_{{\bf Q}_n + q \, \sigma} \\ \Delta_{{\bf Q}_n + q \, \sigma}^* & \left[- v_F p - \sigma h\right] \delta_{q,0} \end{matrix}\right) \left(\begin{matrix}\psi_{R\sigma}(p + q) \\ \psi_{L\sigma}(p)\end{matrix}\right) \nonumber \\
&\qquad \qquad \qquad - \frac{1}{2 N} \sum_{\sigma, \sigma'} \sum_q \, V_{\sigma \sigma'} \left(D_{{\bf Q}_n + q \, \sigma}^* D_{{\bf Q}_n + q \, \sigma'} + D_{{\bf Q}_n + q \, \sigma} D_{{\bf Q}_n + q \, \sigma'}^*\right) .
\end{align}
\end{widetext}
We transform the spinors according to
\begin{equation}
\left(\begin{matrix}\psi_{R\uparrow}(p) \\ \psi_{L\uparrow}(p) \\ \psi_{R\downarrow}(p) \\ \psi_{L\downarrow}(p)\end{matrix}\right) \rightarrow  \left(\begin{matrix}\psi_{R\uparrow}(p) \\ \psi_{L\uparrow}(p) \\ \psi_{L\downarrow}^\dagger(- p) \\ \psi_{R\downarrow}^\dagger(- p)\end{matrix}\right) .
\end{equation}
This transformation maps $D_{{\bf Q}_n + q \, \sigma} \to \sigma D_{{\bf Q}_n + q \, \sigma}$ and the \emph{c}-number term of Eq.~(\ref{v7}) becomes
\begin{equation}\label{va8}
- \frac{1}{N} \sum_{\sigma, \sigma'} \sum_q \, \left(U_c \, \sigma \sigma' - U_s\right) D_{{\bf Q}_n + q \, \sigma}^* D_{{\bf Q}_n + q \, \sigma'} .
\end{equation}
The condensate becomes
\begin{align}
\Delta_{{\bf Q}_n + q \, \uparrow} &\to S(q) - {\rm i} P(q) \quad {\rm and} \nonumber \\
\Delta_{{\bf Q}_n + q \, \downarrow} &\to S(q) + {\rm i} P(q) \label{Deltadown} ,
\end{align}
where we define
\begin{align}
S(q) &= U_c \left(D_{{\bf Q}_n + q \, \uparrow} - D_{{\bf Q}_n + q \, \downarrow}\right) \nonumber \\
&= \frac{U_c}{2} \sum_\sigma \sigma \left(D_{{\bf Q}_n + q \, \sigma} - D_{- {\bf Q}_n + q \, \sigma}\right) \quad {\rm and} \\
{\rm i} P(q) &= U_s \left(D_{{\bf Q}_n + q \, \uparrow} + D_{{\bf Q}_n + q \, \downarrow}\right) \nonumber \\
&= \frac{U_s}{2} \sum_\sigma \left(D_{{\bf Q}_n + q \, \sigma} + D_{- {\bf Q}_n + q \, \sigma}\right) .
\end{align}
The particle hole symmetry (\ref{v6}) implies the relations
\begin{equation}
S(q)^* = S(-q) \qquad ({\rm i} P(q))^* = - {\rm i} P(-q) 
\end{equation}
and we can rewrite Eqs.~(\ref{v7}) and (\ref{va8}) enclosing the system in a box with length $L$ in $x$ direction:
\begin{align}
\frac{H}{N} &= \sum_\sigma \sum_{p, q} \left(\psi_{R\sigma}^{\dagger}(p + q) \ \psi_{L\sigma}^{\dagger}(p)\right) \nonumber \\
&\quad \times \left(\begin{matrix} \left[v_F (p + q) - h\right] \delta_{q,0} & \sigma S(q) - {\rm i} P(q) \\ \sigma S(-q) + {\rm i} P(-q) & \left[- v_F p - h\right] \delta_{q,0} \end{matrix}\right) \nonumber \\
&\quad \times \left(\begin{matrix}\psi_{R\sigma}^{}(p + q) \\ \psi_{L\sigma}^{}(p)\end{matrix}\right) - \sum_q \frac{L |S(q)|^2}{U_c L N} + \sum_q \frac{L |P(q)|^2}{U_s L N} .
\end{align}
Finally, if we transform the spin-$\downarrow$ spinors according to
\begin{equation}
\left(\begin{matrix}\psi_{R\downarrow}(p) \\ \psi_{L\downarrow}(p)\end{matrix}\right) \rightarrow  \left(\begin{matrix}- \psi_{L\downarrow}(p) \\ \psi_{R\downarrow}^\dagger(p)\end{matrix}\right) ,
\end{equation}
we can recast the theory in the form of the genGN model in a chiral basis with coupling constants $g^2 N = - U_c L N/2 v_F$ and $G^2 N = U_s L N/2 v_F$:
\begin{align}
\frac{H}{N} &= \sum_\sigma \int {\rm d}x \, \psi^\dagger(x) \left[- {\rm i} \gamma^5 \partial_x - h + S(x) \gamma^0 \right. \nonumber \\
&\quad + \left. {\rm i} P(x) \gamma^1\right] \psi(x) + \frac{S^2(x)/v_F}{2 (- U_c L N/2 v_F)} \nonumber \\
&\quad + \frac{P^2(x)/v_F}{2 (U_s L N/2 v_F)} ,
\end{align}
where we redefine $S/P(q) \to L^{1/2} S/P(q)$ in order to give proper engineering dimension. The additional spin degree of freedom results in a doubling of flavors in the genGN model. The transformed condensates are
\begin{align}
S(q) &= \frac{U_c}{2} \sum_\sigma \left(D_{{\bf Q}_n + q \, \sigma} + D_{- {\bf Q}_n + q \, \sigma}\right) \nonumber \\
&= \frac{U_c}{2} \sum_\sigma \sum_p \langle \overline{\psi}(p - q) \, \psi(p) \rangle \quad {\rm and} \\
{\rm i} P(q) &= \frac{U_s}{2} \sum_\sigma \left(D_{{\bf Q}_n + q \, \sigma} - D_{- {\bf Q}_n + q \, \sigma}\right) \nonumber \\
&= - \frac{U_s}{2} \sum_\sigma \sum_p \langle \overline{\psi}(p - q) {\rm i} \gamma^5 \, \psi(p) \rangle ,
\end{align}
as one would expect from varying the grand canonical potential density $\Omega = -  \ln {\rm tr} \, e^{- \beta H} /\beta L$ with respect to $S$ and $P$.

As for BCS theory, the cutoff is a physical quantity. The cutoff dependence of the second order transition line between massive and massless homogeneous phases has already been determined in Sec.~III. As pointed out in \cite{2}, the genGN condensate can be approximated by the variational ansatz
\begin{equation}
S(x) = 2 S_1 \cos(2 p_F x) \ {\rm and} \ P(x) = 2 P_1 \cos(2 p_F x)
\end{equation}
in the vicinity of the second order transition between inhomogeneous and homogeneous massless phase. The single particle energies can be calculated perturbatively using almost degenerate perturbation theory which allows us to determine the correction to the grand canonical potential of the free Fermi gas:
\begin{align}
\Psi &= \Psi_{\rm normal} + \delta \Psi \nonumber \\
&= \Psi_{\rm normal} + {\cal M}_{11} S_1^2 + 2 {\cal M}_{12} S_1 P_1 + {\cal M}_{22} P_1^2 .
\end{align}
Varying this expression with respect to the parameters $S_1, P_1$ and $p_F$ yields two conditions which determine the critical values at the phase transition,
\begin{align}
\det {\cal M} &= {\cal M}_{11} {\cal M}_{22} - {\cal M}_{12}^2 = 0 \quad {\rm and} \quad \frac{\partial \det {\cal M}}{\partial p_F} = 0 ,
\end{align}
where the coefficients ${\cal M}_{ij}$ are:
\begin{figure}[b!]
\begin{center}
\scalebox{0.68}{\epsfig{file=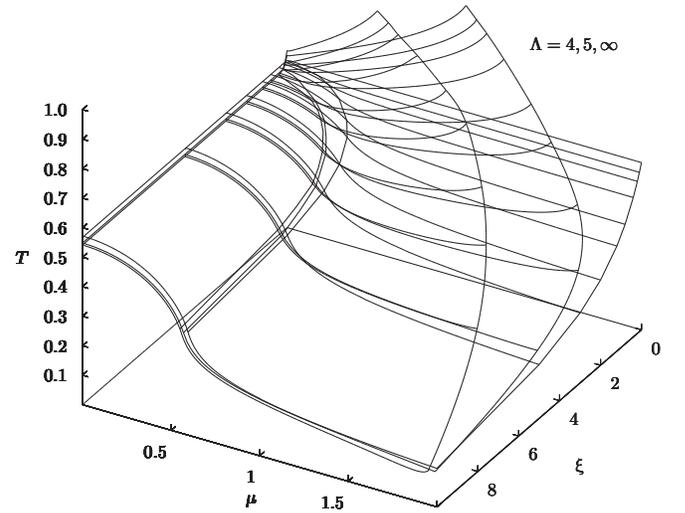}}
\caption{Second order phase transition sheet of the genGN model for the cutoff values $\Lambda = 4,5$ and $\infty$.} 
\label{fig5}
\end{center}
\end{figure}
\begin{align}
{\cal M}_{11} &= {\rm PV} \int_0^{\Lambda/2} \frac{{\rm d}p}{2 \pi} \frac{4 p_F}{p^2 - p_F^2} \left(\frac{1}{1 + e^{\beta (p - \mu)}} + \frac{1}{1 + e^{\beta (p + \mu)}}\right) \nonumber \label{v8} \\
&\quad + \frac{1}{\pi} \ln \left[\frac{\Lambda}{2} + \left(\left(\frac{\Lambda}{2}\right)^2 + 1\right)^{1/2}\right] \nonumber \\
&\quad - \frac{1}{2 \pi} \ln\left[ \left(\frac{\Lambda}{2 p_F}\right)^2 - 1 \right] \\
{\cal M}_{22} &= {\cal M}_{11} + \frac{\xi}{\pi} \\
{\cal M}_{12} &= {\rm PV} \int_0^{\Lambda/2} \frac{{\rm d}p}{2 \pi} \frac{2 p_F}{p^2 - p_F^2} \left(\frac{1}{1 + e^{\beta (p - \mu)}} - \frac{1}{1 + e^{\beta (p + \mu)}}\right) \nonumber \\
&\quad + \frac{1}{2 \pi} \left(\ln\left[\frac{\Lambda}{2 p_F} - 1\right] - \ln\left[\frac{\Lambda}{2 p_F} + 1\right]\right) \label{v9} .
\end{align}
\noindent ${\rm PV}$ denotes a principal value integration. As $\Lambda \to \infty$ the second lines of Eq.~(\ref{v8}) equals $- 1/\pi \ln 2 p_F$. The second line of Eq.~(\ref{v9}) vanishes in this limit and Eqs.~(\ref{v8})-(\ref{v9}) are, of course, equivalent to Eq.~(129) of \cite{2}. Examples of second order transition sheets are shown in Fig.~\ref{fig5}. As expected, even for moderate values of $\Lambda$ the distortion of the phase diagram is negligible. We refrain from determining the cutoff dependence of the first oder transition sheet which would require an extensive numerical Hartree-Fock calculation \cite{47,48} and would not yield much physical insight.

\begin{figure}[t!]
\begin{center}
\scalebox{1.05}{\epsfig{file=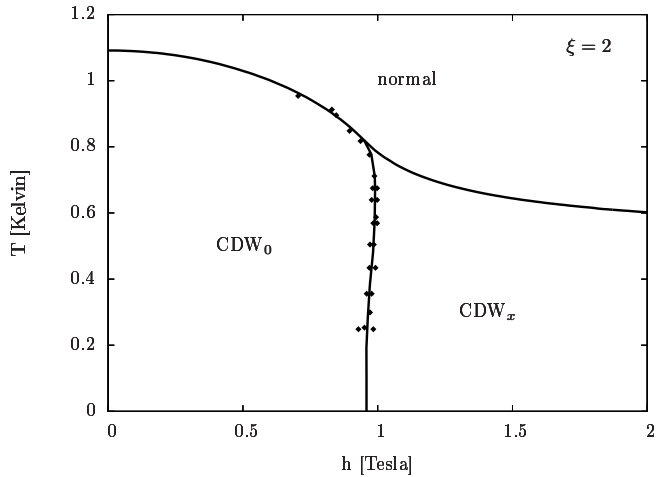}}
\caption{Phase diagram of the inorganic spin-Peierls cuprate CuGeO${}_3$. The fit is done with $\xi \approx 2$. The theoretical curve is fit to the scale of the data. The data is taken from \cite{46}.} 
\label{fig4}
\end{center}
\end{figure}

The phase diagram of various spin-Peierls compounds has been measured with high accuracy \cite{43,44,45,46} whereas theoretical investigations \cite{39,41,42} have not revealed the full phase diagram to the best of our knowledge. In particular, the first order transition line between massive homogeneous (CDW${}_0$) and inhomogeneous (CDW${}_x$) phase has never been determined. We are now able to exploit our mapping and confront experimental data with a full theoretical phase diagram for the first time. Figure~(\ref{fig4}) shows a fit to data obtained by Hase \emph{et al}. \cite{46} for the inorganic spin-Peierls system CuGeO${}_3$~. The theoretical phase diagram is fit to the scale of experimental data. This corresponds to fitting the scale parameters $\Delta_0$ and $v_F$ which are set equal to $1$ in the analysis of the genGN model. The phase diagram of the genGN model was determined in the $\mu$-$T$ plane for fixed values of $\xi = 0, 0.1, 0.2, 0.4, 0.8, 1.2, 2, 3, 5$ and $10$ and the fit $\xi = 2$ was chosen from this ensemble.

\section{Summary and conclusions}

In this work we explored applications of the Gross--Neveu model to nonrelativistic field theories. Starting from the striking observation that the phase diagram of BCS theory and Gross--Neveu model coincide when restricting to phases with translational invariance, we were able to map BCS theory onto the massless Gross--Neveu model with discrete chiral symmetry on a mean-field level. We were able to show that the mean-field Hamiltonian of the quasi one-dimensional extended Hubbard model, which is widely used in condensed matter physics in the description of spin-Peierls systems, is equivalent to a generalized Gross--Neveu model with two coupling constants. In particular, the phase diagrams of both models are equivalent including inhomogeneous phases. This model was worked out in detail only recently by Boehmer and Thies \cite{2}. Relying on their results we were able to complete the phase diagram of the Hubbard model and confront experimental data with the full phase diagram for the first time.

It is interesting to note that although all models have been subject to intensive research over the last three decades the correspondence of the phase diagrams has not been noticed. In particular inhomogeneous phases of the Hubbard model had been discussed long before the first phase diagram of the Gross--Neveu model was even proposed \cite{41,18}.

This work supplements current efforts to use the expertise on quantum field theoretical toy models in the study of phases of QCD at high density by providing systems that can be dimensionally reduced most clearly. The equivalence of the Gross--Neveu and Hubbard model might lay the ground for further work. In particular, analytical solutions for phases of the Gross--Neveu model with continuous chiral symmetry that do not affect the phase diagram might find a physical counterpart \cite{54,55}. Recently, there has been a lot of interest in baryon scattering and other dynamical phenomena in the Gross--Neveu model \cite{52,53}. The extensive analytical work over the past decades that was devoted to the study of dynamical phenomena in spin-Peierls systems (''sliding'' of CDWs) \cite{51} might be useful.

\section*{Acknowledgements}

I would like to thank Michael Thies for suggesting to work out a correspondence between the massless Gross--Neveu model and BCS theory. Furthermore, I am indebted to him for many useful discussions and a critical reading of this manuscript. I would like to thank Christian Boehmer and Dominik Nickel for helpful comments concerning their work and Gerald V. Dunne for pointing out Ref.~\cite{44} and a helpful comment.


\begin{thebibliography}{99}
\bibitem{1}
D. J. Gross and A. Neveu, Phys. Rev. D {\bf 10}, 3235 (1974).
\bibitem{33}
Y. Nambu and G. Jona-Lasinio, Phys. Rev. {\bf 122}, 345 (1961).
\bibitem{18}
U. Wolff, Phys. Lett. B {\bf 157}, 303 (1985).
\bibitem{32}
A. Barducci, R. Casalbuoni, M. Modugno, G. Pettini, and R. Gatto, Phys. Rev. D {\bf 51}, 3042 (1995).
\bibitem{11}
V. Sch\"on and M. Thies, Phys. Rev. D {\bf 62}, 096002 (2000).
\bibitem{58}
M. Thies, Phys. Rev. D {\bf 69}, 067703 (2004).
\bibitem{59}
O. Schnetz, M. Thies, and K. Urlichs, Annals of Physics {\bf 314}, 425 (2004).
\bibitem{60}
O. Schnetz, M. Thies, and K. Urlichs, Annals of Physics {\bf 321}, 2604 (2006).
\bibitem{61}
M. Thies, J. of Phys. A {\bf 39}, 12707 (2006).
\bibitem{55}
G. Basar and G. V. Dunne, Phys. Rev. Lett. {\bf 100}, 200404 (2008).
\bibitem{35}
R. E. Peierls, \emph{More Surprises In Theoretical Physics} (Princeton University Press, Princeton, 1991), Chap. 2.3.
\bibitem{30}
G. Basar, G. V. Dunne, and D. E. Kharzeev, Phys. Rev. Lett. {\bf 104}, 232301 (2010).
\bibitem{34}
E. Shuster and D. T. Son, Nucl. Phys. {\bf B573}, 434 (2000).
\bibitem{20}
T. Kojo, Y. Hidaka, L. McLerran, and R. D. Pisarski, Nucl. Phys. {\bf A843}, 37 (2010).
\bibitem{27}
D. V. Deryagin, D. Yu. Grigoriev, and V. A. Rubakov, Int. J. of Mod. Phys. A {\bf 7}, 659 (1990).
\bibitem{21}
L. McLerran and R. D. Pisarski, Nucl. Phys. {\bf A796}, 83 (2007).
\bibitem{20a}
T. Kojo, R. D. Pisarski, and A. M. Tsvelik, arXiv/1007.0248v1.
\bibitem{1b}
D. Nickel, Phys. Rev. D {\bf 80}, 074025 (2009).
\bibitem{28}
W. Bietenholz, A. Gfeller, and U. J. Wiese, J. High Energy Phys. 10 (2003) 018.
\bibitem{1a}
K. G. Klimenko, Theor. Math. Phys. {\bf 66}, 252 (1986); ibid. {\bf 70}, 125 (1987).
\bibitem{2}
C. Boehmer and M. Thies, Phys. Rev. D {\bf 80}, 125038 (2009).
\bibitem{25}
S. Coleman, Comm. Math. Phys. {\bf 31}, 259 (1973).
\bibitem{31}
E. Witten, Nucl. Phys. {\bf B145}, 110  (1978).
\bibitem{29}
L. Dolan and R. Jackiw, Phys. Rev. D {\bf 9}, 3320 (1974).
\bibitem{3}
G. 't~Hooft, Nucl. Phys. {\bf B72}, 461 (1974).
\bibitem{4}
R. F. Dashen, B. Hasslacher, and A. Neveu, Phys. Rev. D {\bf 12}, 2443 (1975).
\bibitem{4a}
R. Pausch, M. Thies, and V. L. Dolman, Z. Phys. A 338, 441 (1991).
\bibitem{19}
G. Sarma, J. Phys. Chem. Solids {\bf 24}, 1029 (1963).
\bibitem{26}
J. Bardeen, L. N. Cooper and J. R. Schrieffer, Phys. Rev. {\bf 108}, 1175 (1957).
\bibitem{24}
J. Polchinski, arXiv:hep-th/9210046.
\bibitem{61a}
R. Casalbuoni and G. Nardulli, Rev. Mod. Phys. {\bf 76}, 263 (2004).
\bibitem{62}
L.D. Landau and E.M. Lifshitz, \emph{Course of Theoretical Physics}, Statistical Physics Vol. 5 (Butterworth-Heinemann, Washington , DC 1980), \S 57.
\bibitem{22}
R. Shankar, Rev. Mod. Phys. {\bf 66}, 129 (1994).
\bibitem{48a}
H. Ibach and H. L\"uth, \emph{Solid-State Physics} (Springer, New York, 2009), Chap. 10.6, Tab. 10.1.
\bibitem{49}
P. Fulde and R. A. Ferrell, Phys. Rev. {\bf 135}, A550 (1964).
\bibitem{50}
A. I. Larkin and Yu. N. Ovchinnikov, Zh. Eksp. Teor. Fiz. {\bf 47}, 1136 (1964) [Sov. Phys. JETP {\bf 20}, 762 (1965)].
\bibitem{44}
J. P. Boucher and L. P. Regnault, J. Phys. I France {\bf 6}, 1939 (1996).
\bibitem{63}
J. S{\'o}lyom, Adv. in Physics {\bf 28}, 201 (1979).
\bibitem{39}
D. Zanchi, A. Bjelis, and G. Montambaux, Phys. Rev. B {\bf 53}, 1240 (1996).
\bibitem{38}
J. Hubbard, Proc. R. Soc. Lond. A {\bf 276}, 238 (1963).
\bibitem{23}
V. M. Filev, Teor. i Mat. Fiz. {\bf 33}, 119 (1977) [Theor. and Math. Phys. {\bf 33}, 918 (1977)].
\bibitem{36}
E. Melzer, Nucl. Phys. {\bf B443}, 553 (1995).
\bibitem{37}
F. Woynarovich and P. Forg\'acs, Nucl. Phys. {\bf B498}, 565 (1997); {\bf B538}, 701 (1999).
\bibitem{37a}
G.-S. Tian, Phys. Lett. A {\bf 228}, 383 (1997).
\bibitem{47}
C. Boehmer, U. Fritsch, S. Kraus, and M. Thies, Phys. Rev. D {\bf 78}, 065043 (2008).
\bibitem{48}
C. Boehmer, F. Karbstein, and M. Thies, Phys. Rev. D {\bf 77}, 125031 (2008).
\bibitem{43}
C. Proust, A. Audouard, A. Kovalev, D. Vignolles, M. Kartsovnik, L. Brossard,  and N. Kushch, Phys. Rev. B {\bf 62}, 2388 (2000).
\bibitem{45}
J. A. Northby, H. A. Groenendijk, and L. J. de Jongh, J. C. Bonner, I. S. Jacobs and L. V. Interrante, Phys. Rev. B {\bf 25}, 3215 (1982).
\bibitem{46}
M. Hase, I. Terasaki, K. Uchinokura, M. Tokunaga, N. Miura, and H. Obara, Phys. Rev. B {\bf 48}, 9616 (1993).
\bibitem{42}
P. D. Grigoriev and D. S. Lyubshin, Phys. Rev. B {\bf 72}, 195106 (2005).
\bibitem{41}
M. C. Cross, Phys. Rev. B {\bf 20}, 4606 (1979).
\bibitem{54}
G. Basar, G. V. Dunne, and M. Thies, Phys. Rev. D {\bf 79}, 105012 (2009).
\bibitem{52}
W. Brendel and M. Thies, Phys. Rev. D {\bf 81}, 085002 (2010).
\bibitem{53}
A. Klotzek and M. Thies, J. of Phys. A {\bf 43}, 375401 (2010).
\bibitem{51}
Yu. A. Kivshar and B. A. Malomed, Rev. Mod. Phys. {\bf 61}, 763 (1989).

\end{thebibliography}
\end{document}